\documentclass[aps,twocolumn]{revtex4}

\usepackage{amsfonts,color,bbm,amsmath,amssymb}
\usepackage{graphicx,calc,epsfig}

\definecolor{blue}{rgb}{0,0,1}
\definecolor{green}{rgb}{0,1,0}
\definecolor{red}{rgb}{1,0,0}
\definecolor{van}{rgb}{1,0,1}
\definecolor{al}{rgb}{1,1,0}

\newcounter{mnotecount}[section]

\newcommand{\Z}{\mathbb{Z}}

\newcommand{\be}{\nopagebreak[3]\begin{equation}}
\newcommand{\ee}{\end{equation}}
\newcommand{\ba}{\nopagebreak[3]\begin{eqnarray}}
\newcommand{\ea}{\end{eqnarray}}
\DeclareFontFamily{U}{rsfs}{}         % Formal Script            %
\DeclareFontShape{U}{rsfs}{m}{n}{<5> rsfs5 <6><7> rsfs7          %
  <8><9><10><10.95><12><14.4><17.28><20.74><24.88> rsfs10}{}     %
\DeclareMathAlphabet{\mathfs}{U}{rsfs}{m}{n}                     %
\newcommand{\mfs}[1]{\mathfs {#1}}                               %
\newcommand{\va}{\scriptscriptstyle}
\newcommand{\Ag}{{}^{\va \gamma}\!\!A}
\newcommand{\Pg}{{}^{\va \gamma}\! P}
\newcommand{\Pgt}{{}^{\va \gamma\theta}\!P}

\newcommand{\sH}{{\mfs H}}

\newcommand{\sM}{{\mfs M}}
\newcommand{\sN}{{\mfs N}}

\newcommand{\sG}{{\mfs G}}

\newcommand{\Cyl}{{Cyl}}

\usepackage{psfrag}
\usepackage{vmargin}
\usepackage{amsfonts}
\usepackage{amsmath}
\usepackage{graphics,graphicx,color,calc,epsfig}
\usepackage{euscript}
\newcommand{\beq}{\begin{equation}}
\newcommand{\eeq}{\end{equation}}
\newcommand{\beqa}{\begin{eqnarray}}
\newcommand{\eeqa}{\end{eqnarray}}

\newcommand{\R}{\mathbb{R}}

\newcommand{\tmop}[1]{\operatorname{#1}}

\begin{document}

\title{The $\theta$ parameter in loop quantum gravity: effects on quantum geometry and black hole entropy}

\author{Danilo Jimenez Rezende}
\affiliation{Centre de Physique Th\'eorique\footnote{Unit\'e Mixte
de Recherche (UMR 6207) du CNRS et des Universit\'es Aix-Marseille
I, Aix-Marseille II, et du Sud Toulon-Var; laboratoire afili\'e
\`a la FRUMAM (FR 2291)}, Campus de Luminy, 13288 Marseille,
France.}

\author{Alejandro Perez}
\affiliation{Centre de Physique Th\'eorique\footnote{Unit\'e Mixte
de Recherche (UMR 6207) du CNRS et des Universit\'es Aix-Marseille
I, Aix-Marseille II, et du Sud Toulon-Var; laboratoire afili\'e
\`a la FRUMAM (FR 2291)}, Campus de Luminy, 13288 Marseille,
France.}

\date{\today \vbox{\vskip 2em}}

\begin{abstract}
The precise analog of the $\theta$-quantization ambiguity of
Yang-Mills theory exists for the real $SU(2)$ connection
formulation of general relativity. As in the former case $\theta$
labels representations of large gauge transformations, which are
super-selection sectors in loop quantum gravity. We show that
unless $\theta=0$, the (kinematical) geometric operators such as
area and volume are not well defined on spin network states. More
precisely the intersection of their domain with the dense set
$Cyl$ in the kinematical Hilbert space $\sH$ of loop quantum
gravity is empty. The absence of a well defined notion of area
operator acting on spin network states seems at first in conflict
with the expected finite black hole entropy. However, we show that
the black hole (isolated) horizon area---which in contrast to
kinematical area is a (Dirac) physical observable---is indeed well
defined, and quantized so that the black hole entropy is
proportional to the area. The effect of $\theta$ is negligible in
the semiclassical limit where proportionality to area holds.

\end{abstract}

\maketitle

%%%%%%%%%%%%%%%%%%%%%%%%%%%%%%%%%%%%%%%%%%%%%%%%%%%%%%%%%%%%%%%%%%%%%
\section{Introduction}\label{int}
%%%%%%%%%%%%%%%%%%%%%%%%%%%%%%%%%%%%%%%%%%%%%%%%%%%%%%%%%%%%%%%%%%%%%

A remarkable feature of general relativity (GR) is that it admits
a connection formulation with a (unconstrained) phase space
isomorphic to that of $SU(2)$ Yang Mills theory \cite{AA}. This property of
GR is of great importance for the definition of the quantization
program of loop quantum gravity (LQG). At the basic level, LQG is
defined using non perturbative techniques first developed in the
context of Yang-Mills theories. For instance, the use of the
parallel transport of the $SU(2)$ connection as configuration
variable---which combined with diffeomorphism invariance---allows
for the precise definition of a (background independent) approach
to quantum gravity. The compactness of the associated gauge group
lies at the heart of the very existence of the representation of
the fundamental operators in a Hilbert space where the constraints
of GR can be promoted to finite operators (see \cite{lqg} and
refs. therein).

The $SU(2)$ connection  formulation of GR is defined in terms of
the so-called Ashtekar-Barbero variables which are labelled by a
real parameter $\gamma$ known as the Immirzi parameter. Here we
shall see that a more general $SU(2)$ connection formulation of GR
includes a new dimensionless parameter $\theta\in [0,2\pi]$. These
two-parameter family of descriptions of GR are all classically
equivalent. However, upon quantization the pair $(\gamma,\theta)$
labels unitarily inequivalent theories and therefore represent a
quantization ambiguity of LQG.

In the following section we
show how the two parameter family of $SU(2)$ connection
formulations can be obtained from a series of canonical
transformations starting from the standard ADM or metric
Hamiltonian formulation of GR. This implies that all these
formulations are classically equivalent.

In Section \ref{lgt} we show how the newly introduced $\theta$
parameter labels unitary representations of large $SU(2)$ gauge
transformations which are super-selected sectors of quantum
gravity.

In Section \ref{qg} we study the implications of the $\theta$
ambiguity for the definition of (kinematical) quantum geometric
operators such as area and volume. We show that the action of the
latter is not well defined on spin network states. We conjecture
that their domain is not dense in the Hilbert space $\sH$ of LQG.
This might seem problematic if one would like to attribute a
physical meaning to kinematical quantities. Since, in LQG the
quantization of the area operator plays an important role in the
computation of black hole entropy \cite{Ashtekar:2000hw,bhc}, one
might expect at first sight difficulties with the computation of
black hole entropy for $\theta\not=0$.

However, in contrast with kinematical area, the black hole horizon
area (which is a true observable in the classical theory) can be
quantized and has a discrete spectrum in the $\theta\not=0$. In
Section \ref{bhe} we show how the standard result for the black
hole entropy is recovered in the semiclassical limit. In Section
\ref{disc} we interpret this result.

\section{More general connection formulation of GR}\label{nv}

The $SU(2)$ Ashtekar-Barbero variables can be introduced in
several ways. Perhaps the shortest path is the one defined by a
series of canonical transformations starting from the Arnowitt,
Deser, and Misner (ADM) variables for general relativity. The ADM
parametrization of the (unconstrained) phase space of gravity is
given by the canonical pair $(q_{ab},P^{ab})$ with $a,b=1,2,3$
space indices. The configuration variable $q_{ab}$ is the
$3$-metric of a Cauchy surface $\Sigma$ (the space time manifold
$\sM$ is assumed to be of topology $\sM=\Sigma\times \R$), while
the momentum $P^{ab}$ is defined in terms of the extrinsic
curvature $K_{ab}$ (Lie derivative of $q_{ab}$ in the direction
normal to $\Sigma$) as
\[P^{ab}=q^{-1/2}(K^{ab}-K q^{ab}),\]
where $q^{ab}$ is the inverse of $q_{ab}$, $q=\det(q_{ab})$, and
$K=q_{ab}K^{ab}$.

One can introduce a (densitized) triad defined by the equation \be
q q^{ab}=E_j^aE_i^b \delta^{ij}.\label{uno}\ee With $E^a_i$ at hand one
introduces the quantity \be
K_a^i=q^{-1/2}K_{ab}E^b_j\delta^{ij}.\ee It is easy to show
that---provided $E^a_i$ and $K_b^j$ satisfy the
constraint \footnote{This constraint comes from the fact that
$K_{ab}=K_{ba}$ and can be shown to generate local $SU(2)$
rotations of the triad canonical variables.} \be\label{gg}
G_i=\epsilon_{ijk}E^{aj}K_a^k\approx 0\ee---one can express the ADM
variables in terms of the pair $E^a_i$ and $K_a^i$, and that the
latter are indeed canonically conjugate variables.

The Ashtekar-Barbero variables are obtained by the canonical
transformation \be \label{barbero}\Pg^a_i=\gamma^{-1}E^a_i\ \ \ \
\Ag_a^i=\Gamma_a^i(E)+\gamma K_a^i,\ee where $\Gamma_a^i$ is the
torsion free spin connection (i.e. a functional of $E^a_i$ alone),
and $\gamma\in \R$ is an arbitrary parameter known as the Immirzi
parameter. It is easy to see that the previous transformation is a
canonical transformation \cite{titi} if one realizes that
$\Gamma_a^i={\delta}W_1[E]/{\delta E^a_i}$ with \be
W_1[E]=\int_{\Sigma} \epsilon_{bcd}E^a_{[i}E^b_{j]}\partial_a
\frac{E^{ci}E^{dj}}{\det(E)}.\ee In terms of the new variables the
constraint (\ref{gg}) becomes the Gauss law of $SU(2)$ Yang-Mills
theory, namely \be\label{gauss} G_i=D_a\Pg^a_i\approx 0,\ee where
$D_a$ is the covariant derivative defined in terms of the
connection $\Ag_a^i$.

Are there more general connection variables than the ones obtained
above? Yes, if we are given a background independent functional
$W_2[\Ag]$, since
 one can leave $\Ag$ unchanged and define a new momentum variable
$(\Pgt^a_i)=\Pg^a_i+\delta
W_2[\Ag]/\delta \Ag_a^i$. In three dimensions there is only a one parameter family of $W_2[\Ag]$
%\footnote{One needs the generating functionals to be diff
%invariant so that the functional is diff covariant, i.e. a tensor
%in the spacetime indices.}.
given by \be
\label{w2}W_2=\frac {\theta}{16\pi^2} \int_{\Sigma}{\rm Tr}[\Ag\wedge
d\Ag+\frac{2}{3} \Ag\wedge \Ag\wedge \Ag],\ee where $\theta$ is a
real parameter and the integral is the well known Chern-Simons
action. In terms of the canonical pair $(E_a^i,K_b^j)$ we
get\ba\label{new} \Pgt^a_i=\gamma^{-1} E^a_{i} +\frac {\theta}{8\pi^2}\, B^a_i \
\ \ \Ag_a^i=\Gamma_a^i+ \gamma K_a^i,\ea where
$B^a_i=\epsilon^{abc}F_{bc}^j\delta_{ij}$ is the non Abelian
magnetic field of $\Ag_a^i$. Due to the Bianchi identity the Gauss
constraint has the same functional form as (\ref{gauss}) where
$\Pg$ is replaced by $\Pgt$. From the remaining constraints only
the Hamiltonian or scalar constraint has an explicit
$\theta$-dependent P violating term \cite{Ashtekar:1988sw} (see
also \cite{Montesinos:2001ww}).

\section{Large $SU(2)$ gauge transformations}\label{lgt}

There is a nice geometric way of understanding the previous
quantization ambiguity \cite{Ashtekar:1988sw} (see also
\cite{Mercuri:2007ki} and  \cite{Gambini:1998ip} for a similar argument and analogies concerning the origin
of $\gamma$). In the quantum theory states are required to be
anihilated by the quantum Gauss constraint. As the latter is the
infinitesimal generator of local $SU(2)$ transformations, states
are required to be invariant under the gauge group $\sG_{\va 0}$
of transformations connected to the identity. However, assuming
for the moment that $\Sigma$ is compact, and due to the fact that
$\pi_3(SU(2))=\Z$, the full gauge group of gravity $\sG$ also
contains the so-called {\em large gauge transformations}. The
latter are gauge transformation $g(x)$ with non trivial winding
number \be\nonumber w[g]=\frac{1}{24 \pi^2}\int {\rm tr}[g^{-1}dg
\wedge g^{-1}dg\wedge g^{-1}dg].  \ee Now if $\alpha\in \sG$ is
such that $w[\alpha]=1$ then one can show that $\sG/\sG_{\va
0}\approx\{\alpha^n/ n\in \Z\}$. Therefore, physical states can
transform in a non trivial fashion under large gauge
transformations. We denote $\sH_{\theta}$ the unitary
(irreducible) representations of $\sG/\sG_{\va 0}\approx \Z$ which
are labelled by an angle $\theta\in [0,2\pi]$. Given $\Psi\in
\sH_{\theta}$ and $\alpha\in \sG$ (s.t. $w[\alpha]=1$)\be \alpha
\triangleright \Psi = e^{i\theta} \Psi.\ee Finally due to
the fact that physical observables are invariant under the full
group $\sG$ they leave $\sH_{\theta}$ invariant. Hence, each
$\sH_{\theta}$ gives a different quantization.

The relationship with the analysis of the previous section is
clarified if one realizes that the non trivial transformation rule
for states in $\sH_{\theta}$ can be shifted to operators by means
of the state redefinition \be\label{transf} \tilde \Psi[A]=\exp{(-i
W_2[A])}\Psi[A] \ee with $W_2[A]$ as defined in (\ref{w2})\footnote{Transformation
(\ref{transf}) corresponds to multiplication by the Kodama state. For the relevance of such state for quantum gravity see \cite{kodama}.}. Hence
working with wave functions with non trivial behaviour under large
gauge transformation is equivalent to working with a transformed
momentum\ba\label{vv}\nonumber &&
\Pgt^a_i=\exp{(-iW_2[A])}\Pg^a_i\exp{(iW_2[A])}
%\\ && =\Pg^a_i+2\theta \ B^a_i
,\ea which has the form (\ref{new}) on $\sH_0=\sH$ (spanned by spin network states). Therefore, the $\theta$
ambiguity, described in Section \ref{int}, has a clear meaning in
the quantum theory. It encodes the non trivial representations of
$\sG/\sG_{\va 0}$ which are super selected sectors of quantum
gravity.

\section{Quantum geometry}\label{qg}

In the standard LQG formulation of the quantum theory \cite{lqg}
the fundamental (to-be-quantized) variables are the holonomy of
$\Ag$ along one dimensional paths $e\subset \Sigma$, and fluxes of
$\Pgt$ across surfaces $S\subset \Sigma$. Respectively, \be
h_e[\Ag]={\rm P}e^{{ -}\int\limits_e \Ag} \ \ {\rm and} \ \
\nonumber \Pgt(r,S)=\int\limits_S r\cdot(\epsilon\Pgt), \ee where
the $2$-form $r\cdot \epsilon \Pg_{ab}=r^i \epsilon_{abc}\Pg^c_i$,
and $r^i$ is an arbitrary field of internal directions. These
(kinematical) observables can be represented as self adjoint
operators in the (kinematical) Hilbert space $\sH$ of LQG. The
fundamental excitations in $\sH$ are given by quantized lines of
flux of $\Pgt^a_i$ which can be organized in an orthonormal basis
of (open) spin network states. The quantum flux operators $
\Pgt(r,S)$ have discrete spectrum. More precisely, the basic non
trivial quanta of $\Pgt(r,S)$ are given by a spin network edge
labelled by the spin $j$ and magnetic number $m$ transversal to
$S$ and ending at $S$. In that case the possible values of this
puncture-like contribution to $\Pgt(r,S)$ is $8\pi \ell_p^2 m$ for
$-j\le m\le j$. The general eigenstates are given by states with
$n$ punctures $|n,\{j_i, m_i\}_{i=1}^n \rangle$ for which
\ba\!\!\! (\Pgt(r,S)\! - \!8\pi \ell_p^2\! \sum_{\va i = 1}^{\va n}\!m_i
)\left|n;\{j_i,m_i\}_{i=1}^n \rangle\right.\!\! = 0,\label{nine}
\ea with $-j_i\le m_i\le j_i$. The position of the punctures on
$S$ also labels the previous states. We have simplified the
notation here as the details are not essential for our
argumentation.

There are two simple $SU(2)$ invariant operator that one can
construct (both having discrete spectra \cite{lqg}). The first one
associates to any surface $S\subset \Sigma$ the quantity
\be\nonumber O_1(S,\Pgt)=\int_S \sqrt{(\Pgt^a_i)(\Pgt^b_j)
\delta^{ij}n_a n_b}, \ee where $n_a$ is the normal to the surface
$S$. The other one associates to a three dimensional region
$U\subset\Sigma$ the quantity \be\nonumber O_2(U,\Pgt)=\int_{U}
\sqrt{(\Pgt^a_i)(\Pgt^b_j)(\Pgt^c_k)
\epsilon^{ijk}\epsilon_{abc}}\ee However, according to (\ref{uno})
the field encoding the dynamical Riemannian geometry of $\Sigma$
is \be E^a_i=\gamma (\Pgt^a_i)-\frac {\theta}{8\pi^2}\gamma \
B^a_i(\Ag). \ee Therefore, in the case $\theta=0$ the previous
operators have a clear geometric meaning: $O_1$ relates to the
area of $S$ via ${\cal A}(S)=\gamma O_1(A)$, and $O_2$ relates to
the volume of $U$ via ${\cal V}(U)=\gamma^{\frac{3}{2}} O_2(U,P)$.
However, for $\theta\not=0$, the area and volume depend on both
the connection $\Ag$ (through the magnetic field $B^a_i$) and
$\Pgt$. In that case these operators are not well defined in the
kinematical hilbert space of LQG (see Appendix for an explicit
proof of this statement).

Here, the term {\em kinematical observable} is used to designate
quantities that are not measurable in the theory but serve to
setup Dirac's quantization program. The reason for that is that
the (kinematical) observables introduced above are not invariant
under the full gauge group of GR that includes diffeomorphisms in
addition to local $SU(2)$. This qualification is quite important
for the interpretation of our results. We shall come back to this
in Section \ref{disc} once we elaborate a bit more on the effects
of $\theta$.

\section{Black hole entropy}\label{bhe}

In Section \ref{qg} we argued that the discreteness of kinematic
geometric operators is lost in the $\theta\not=0$ sectors. Despite
of this fact we show here that (due to dynamical effects) the
quantum operator associated to the area of black hole (isolated)
horizon has indeed a discrete spectrum for arbitrary $\theta$.
This leads to a finite entropy of the black hole horizon which is
proportional to the macroscopic area. $\theta$ does not affect the
leading order contribution to the entropy calculation.

The computation of Black Hole entropy is based on the quantization
of a sector of the phase space of GR containing a
so-called isolated horizon (IH) \cite{{Ashtekar:2000hw}} (for
simplicity here we will assume the horizon is non rotating).
Physically a non rotating IH is a three dimensional null surface $\Delta$
equipped with a preferred foliation by $2$-spheres such that the
area of the spheres is constant.

The  boundary conditions on the horizon reduce the $SU(2)$ gauge
symmetries on the bulk to a $U(1)$ subgroup leaving invariant an
internal direction $r^i$. There are non trivial degrees of freedom
at the horizon encoded in the pull back of the bulk connection on the horizon
$H=\Delta\cap \Sigma$; a $U(1)$-connection
$\Ag=\Ag^i r_i$ (notice the obvious abuse of notation). The
validity of Einstein's equations at the horizon imply the
following relationship between bulk and horizon degrees of
freedom: \be\label{const} F_{ab}(\Ag)= - \frac{2 \pi}{a_{\va H}}
{\epsilon_{abc} E^c}_i r^i, \ee where $a_{\va H}$ is the macroscopic area of the horizon.

As shown in \cite{Ashtekar:2000hw} the IH boundary condition
consistently defines a sector of the phase space of gravity. The
requirement that the simplectic structure be foliation independent
introduces a Chern-Simons boundary term. The latter leads to the
quantization of the degrees of freedom on the horizon. More
precisely, the simplectic potential in terms of the variables
$(\Pg,\Ag)$ is
\begin{eqnarray}\label{sp}\nonumber
  && 8 \pi G^{\gamma} \!\Theta ( \delta ) = \\ && -\!\!\! \int_{\Sigma} \tmop{Tr} \delta\Ag
  \wedge \epsilon\Pg  + \frac{a_{H}}{4 \pi \gamma}\!\!\! \int_{H} \
  \delta\Ag \wedge \Ag. \left. \right.
\end{eqnarray}
In order to express it in terms of (\ref{new}) one adds to
$^{\gamma} \!\Theta ( \delta )$ the  `total field differential' of
$W_2[A]$;  \be\nonumber \delta W_2[\Ag] = \theta(2\int_{\Sigma}
\tmop{Tr} F(\Ag) \wedge \delta \Ag + \int_{H} \Ag \wedge \delta
\Ag), \ee and the new potential becomes \ba &&\!\!\!\!\!\!
^{\gamma \theta}\!\! \Theta =
\\ && \nonumber =-\frac{\hbar}{\ell_p^2}\!\int_{\Sigma} \tmop{Tr} \delta\Ag
\wedge\epsilon\Pgt + \kappa\hbar \!\!\!\int_{H} \ \delta\Ag
\wedge\Ag, \ea where $\kappa= a_H/(4 \pi \gamma\ell_p^2) -
{\theta}/{16\pi^2}$. So we see that in addition to the $\theta$
modification (\ref{new}) of the momentum conjugate to $\Ag$ in the
bulk, the canonical transformation has the effect of shifting the
level $\kappa$ of the Chern-Simons contribution to the simplectic
structure. In terms of the phase space variables the (dynamical)
constraint (\ref{const}) takes the simple form \be\label{ee}
  2 \kappa {F} ( \Ag ) = -\epsilon\Pgt^i r_i.
\ee Remarkably the constraint (\ref{ee}) has the same functional
form as the one found in \cite{{Ashtekar:2000hw}} for the
$\theta=0$ case. The $\theta$-dependence is hidden in the explicit
form of the Chern-Simons level $\kappa$ (A similar thing happens
in the quantization of IH in the presence of non minimally coupled
scalar fields \cite{Ashtekar:2003jh}). This implies that the
quantization techniques of \cite{Ashtekar:2000hw} can be directly
applied to the $\theta\not=0$ case. In the quantum theory the
above requirement restricts the set of physical states. The
constraint (\ref{ee}) requires states in the bulk to be
eigen-states of the flux of $\Pgt\cdot r$ across H. These are spin
network edges carrying spins $j$ ending at the horizon such that
they are eigenstates of the r.h.s. of (\ref{ee}). From
(\ref{nine}), the allowed eigenvalues of a single puncture are
$8\pi\ell_p^2 m$ with $-j\le m\le j$. On the other hand, the
quantization of Chern-Simons implies that the holonomy around a
single puncture has eigenvalues $\exp{i2\pi a_i/\kappa}$ with
$a_i\in \Z$ mod $\kappa$. Quantum Einstein eqs. (\ref{ee}) select
those states for which $a_i= - 2 m_i \tmop{mod} \kappa$. Moreover,
states satisfying (\ref{ee}) are eigenstates of
the quantum horizon area. Explicitly one has \ba &&\! \!\!\!\!\!\!\nonumber A_{H} \left|n;\{j_i,m_i\}_{i=1}^n \rangle\right. \\
\nonumber &&\!\!\!\!\!\!\! =8\pi \gamma \ell_p^2 \sum_{\va i =
1}^{\va n}\! \sqrt{ K m_i^2+j_i(j_i+1)}|n,\{j_i, m_i\}_{i=1}^n
\rangle, \ea where $K=\theta(a_H/(4\pi\gamma
\ell_p^2)-{\theta}/{16\pi^2})^{-1}$, $n$ is the number of
punctures, and $\{j_i, m_i\}_{i=1}^n$ is the set of quantum
numbers associated to the punctures on $H$. Notice that
discreteness follows from the quantization of the magnetic flux at
$H$ (compare with the generalized eigenstate defined in the
Appendix). Therefore, here the dynamical condition (\ref{ee})
implies that the quantum operator associated to the Dirac
observable $A_H$ is well defined. Using the counting techniques of
\cite{Meissner:2004ju} one finds that the $\theta$-dependence does
not change the leading term in the entropy: explicitly
$S_H:=\log[\sN(a_H)]\approx (4\ell_p\gamma)^{-1}\gamma_M a_H$,
where $\sN(a_H)$ is the number of horizon states compatible with a
macroscopic horizon area $a_H$ and $\gamma_M=0.23..$.

\section{Discussion}\label{disc}

General relativity admits a two-parameter family of $SU(2)$
connection formulations labelled by the Immirzi parameter
$\gamma\in \R$ and $\theta\in [0,2\pi]$. Our arguments show that
discreteness of (kinematical) geometric operators in LQG is a
special property of the $\theta=0$ sector. For $\theta\not=0$
kinematical area and volume are far more complicated objects. In
the appendix we prove some properties that suggest that they are
not densely defined in $\sH$ (which we conjecture). Nevertheless,
discreteness at the fundamental level remains in the sense that
the (kinematic) Hilbert space is given by the span of spin network
states (labelled graphs and discrete quantum numbers).

What are the implications of this underlying discrete structure?
General covariance implies that only fully gauge invariant
observables (i.e. Dirac observables which are both $SU(2)$ and
diffeomorphism invariant) are physically meaningful. From this
perspective the discreteness of kinematical area and volume
(although an interesting property when present) is not by itself
of direct physical relevance. The physical question is rather
whether the fundamental discreteness of the state space of LQG
would leave imprints in fully gauge invariant quantities which
represent physical observables. Unfortunately, due to the
dynamical nature of these observables, this question is very
difficult to answer in general at this stage of development of
LQG.

Nevertheless, one can try to answer the question in particular
cases. One such case is the IH system, where the black hole area
is a Dirac observable. From our viewpoint this is  the most
important result of the paper: we have shown that due to the
dynamics of general relativity---encoded in the IH boundary
condition---the spectrum of the area of black hole horizon remains
discrete. In this case problems concerning the quantization of
(kinematical) area can be viewed as a gauge artifact that disappear
when looking at gauge invariant IH area.

As in QCD, the angle $\theta$ introduces parity (as well as time
reversal) violation in quantum gravity
\cite{Ashtekar:1988sw}\footnote{Notice that under a parity
transformation the magnetic field $B$ changes sign.}. As a
consequence, one would expect only $P$ violating observables to be
sensitive to $\theta$. The black hole IH system is such an example: notice for instance the $P$-violating nature of the IH boundary condition
(\ref{ee}). Therefore the $\theta$-dependent effects found here
are expected from general considerations. As in the quantum theory
$\theta$ is defined modulo $2\pi$ the latter effects are
sub-leading terms that are not relevant in the semiclassical limit
($a_{\va H}/\ell_p \!\!>\!\!\!> 1$). Our results represent another
non trivial test for the  universality BH entropy in
loop quantum gravity.

As explained in Section \ref{nv} the Gauss and vector constraints
of canonical gravity are unchanged in the $\theta\not=0$ sectors.
However, the scalar constraint now has new P-violating terms. We
have concentrated here on a very specific dynamical situation
where the precise form of the scalar constraint did not play any
important role (in essence the scalar constraint is replaced by
the condition (\ref{ee}) when dealing with IHs). It would be
interesting to investigate the effects of the additional terms to
the dynamics of LQG (of special interest is the case where fermions are present).

The quantization of the (kinematical) volume operator plays a
central role in the quantization of the scalar constraint in the
$\theta=0$ sector \cite{Thiemann}. The difficulties associated
with the quantization of the (kinematical) area operator described
in the appendix would also appear in the quantization of volume in
the $\theta\not=0$ sectors. In this sense, it seems that the usual
quantization techniques applied to the scalar constraint cannot be
imported directly to the generic sectors. This issue  should be
studied in detail.

\section*{Acknowledgements}
This work was supported in part by: the Agence Nationale de la Recherche,
Grant No. ANR-06-BLAN-0050. Thanks to A. Corichi, J. Engle, M. Knecht, J. Lewandowski, M.
Montesinos, J. Pullin, C. Rovelli, H. Sahlmann and
T. Thiemann for discussions.

\begin{appendix}

\section{On the quantization of geometric operators}

Here we argue that the kinematical geometric operators are not
well defined in the $\theta\not=0$ case. We do so by showing
explicitly that the action of the area operator is ill defined on
elements of $Cyl\subset \sH$.

Without loss of generality we can assume that we have local
coordinates $x^1,x^2,x^3$ and that the surface $S$ is defined by
$x^3=0$. In terms of our basic phase space variables the area
$A(S)$ takes the form \ba &&\!\!\!\! \!\!\!\!A(S)=\int \sqrt{E^3\cdot E^3}\\
&&\nonumber \!\!\!\!\!\!\!\!=\gamma\int \sqrt{(P^3-\frac {\theta}{8\pi^2}
B^3)\cdot(P^3-\frac {\theta}{8\pi^2} B^3)}. \ea

In order to quantize the previous expression one needs to
introduce a regularization. We take for example a cellular
decomposition of $S$. As it will become clear below the details of
the regularization do not matter for our argument. Therefore, here
we take the dual of the lattice $(\epsilon n,\epsilon m, 0)$ for
$0< \epsilon\in \R$ and $n,m\in \Z$. Then at the classical level it is easy to
see that \be\label{A2}\! A(S)=\lim_{\epsilon \rightarrow 0} \sum_{n,m}\!
 \sqrt{E(S^{nm},\tau^i)E(S^{nm},\tau_i)},\ee
where $S^{nm}$ is the plaquette dual to the lattice point
$\epsilon (n,m,0)$, and \[E(S^{nm},\tau^i)=\gamma
P(S^{nm},\tau^i)-2\gamma\theta B(S^{nm},\tau^i)\] with
\[B(S,r^i):=\int B^3_ir^i dx^1 dx^2\] and $P(S^{nm},\tau^i)$ defined in Section \ref{qg}.
Let us concentrate for the moment on the action of the regularized
area operator on the constant function $1\in Cyl$. Using the fact that
$P(S^{nm},\tau_i)\triangleright 1=0$ for all $n,m$ we have: \ba
&&\!\!\!\!\!\!\!\!\!\!\!\!\!\!\! \nonumber \lim_{\epsilon \rightarrow
0} \sum_{n,m} \sqrt{E(S^{nm},\tau^i)E(S^{nm},\tau_i)}\triangleright
1=\\ &&\!\!\!\!\!\!\!\!\!\nonumber =\lim_{\epsilon \rightarrow 0}
\sum_{n,m} \sqrt{B(S^{nm},\tau^i)B(S^{nm},\tau_i)}\triangleright 1\\
\nonumber &&=2 \lim_{\epsilon \rightarrow 0} \sum_{n,m} \sqrt{{\rm
Tr}[U^{nm}\tau_i]{\rm Tr}[U^{nm}\tau^i]}\triangleright 1,\ea where
we used that \be\label{uu} B(S^{nm},\tau_i)=2{\rm
Tr}[U^{nm}\tau_i]+{\cal O}(\epsilon^4)\ee in the last line. Notice
that the last line implies that the action on $1\in Cyl$ is given
by a sum of mutually-orthogonal terms in $Cyl\subset \sH$.
Therefor the norm of such regularized action grows with the number
of plaquettes and we have \be ||\sum_{n,m} \sqrt{{\rm
Tr}[U^{nm}\tau_i]{\rm Tr}[U^{nm}\tau^i]}\triangleright 1||\sim
\frac{1}{\epsilon^2}.\ee We conclude that the action of
$A(S)\triangleright 1$ is not defined. From this one can easily
see that the same problem with the limit $\epsilon\rightarrow 0$
arises for any $\phi\in Cyl$: for an arbitrary $\phi\in Cyl$ it
suffices to concentrate on an open region of $S$ that does not
intersect the graph defining $\phi$.  This problem is not
surprising, it is completely equivalent to the one found if one
tries to define for example the Yang-Mills hamiltonian in $\sH$.
The action of the area operator on $Cyl$ is not well defined in
the $\theta\not=0$ sectors.

Is the domain of the area operator dense in $\sH$? It seems that
the only way to avoid the divergences found above one would need
the plaquette actions in the regulated operator to act trivially.
Formally speaking one would need the `quantum' magnetic field to
vanish almost everywhere on $S$. Such states are however outside
of $\sH$ and can only be given a distributional meaning. From this
it seems that it is reasonable to conjecture that the area
operator is not even densely defined in $\sH$. A similar conclusion can
be obtained for the volume.

For completeness we give here an example of distributional states
$\Psi$ in the dual space $Cyl^{\star}$ for which the action of
$A(S)$ is well defined in the sense that $A(s)\triangleright\Psi
\in Cyl^{\star}$ where \be
A(s)\triangleright\Psi(\phi):=\Psi(A(s)\triangleright\phi) \ \ \
\forall \phi\in \Cyl.\ee We shall do this by exhibiting a special
family of states $\Psi_a\in Cyl^{\star}$ labelled by a group
element $a\in SU(2)$. The state is defined by its action on $Cyl$.
It would suffice to define the action of $\Psi_a$ on any element
of the spin network basis.

\noindent {\bf Definition:} For any  $\phi \in Cyl$
we define the action of
$\Psi_a$ by the following three properties:

\begin{enumerate}
\item $\Psi_{a}(1)=1$ \item $\Psi_a(\phi)=0$ if the underlying graph of $\phi$ is not
entirely contained on $S$ (in our coordinates if it does not
lie on the $x^3=0$ plane) \item If the graph of $\phi$ is contained in $S$ then we
decompose $\phi$ in terms of spin networks, and subsequently we write the spin networks as a product
of Wilson loops. In this way we can write any such element of $\phi \in Cyl$ as
\[
\phi=\sum_{L} c_L \prod_{\ell \in L} \alpha_{\ell},
\]
where $L$ are collections of loops, $\ell\in L$ denotes a loop in the collection $L$ and $\alpha_{\ell}$
is the trace of the holonomy in the fundamental representation around $\ell$. The loops $\ell$ are all
contained in $S$, and can
have self-intersections. We define \be \Psi_a(\phi):= \sum_{L} c_L \prod_{\ell \in L} \Psi_a(\alpha_{\ell}),\ee
where $\Psi_a(\alpha_{\ell})={\rm tr}(a^{w[\ell]})$ where $w[\ell]$ is the winding number of $\ell$
around the point $(0,0,0)$.
\end{enumerate}

{\noindent\bf Lemma:} For $S^I\subset S$, any The following identities
follow from the previous definition
\be\label{A1}\Psi_a(P(S^I,\tau^i)P(S^I,\tau_i)\triangleright \phi)=0,\ee and
\ba\label{A2}\nonumber &&\!\!\!\!\!\!\!\!\!\!\!\!\!\!\! \Psi_a(P(S^I,\tau^i)B(S^I,\tau_i)\triangleright
\phi)= \\ && \!\!\!=\Psi_a(B(S^I,\tau_i)P(S^I,\tau^i)\triangleright \phi)=0,\ea
or all $\phi\in Cyl$. \vskip.5cm \noindent {\em Proof:}
If $\phi$ can be expanded in terms
of spin networks fully contained in $S$ then the implication of
the lemma is obvious because $P^3$ commutes with $B^3$ and annihilates such states. In
order to avoid the trivial action of $P^3$ we need spin network
edges that are transversal to the surface $S$. In that case the
action of $P^3$ is non trivial but the resulting state in both (\ref{A1}) and (\ref{A2})
contains edges that are outside
$S$ and therefore the implication of the lemma follows from the
condition (2) of our definition.

\vskip.5cm

\noindent {\bf Proposition:} The (generalized) state $\Psi_a$ is
a (generalized) eigenstate of the area $A(S)$ with eigenvalue
$\gamma \frac {\theta}{4\pi^2}\sqrt{a_ia^i}$, where $a_i={\rm tr}(a\tau_i)$.

\vskip.5cm \noindent {\em Proof:}
We start from the regularized expresion (\ref{A2}) and
concentrate for a moment on the argument
of the square root for the term $(n,m)$. Using the lemma above we have
\ba \nonumber &&\!\!\!\!\!\!\!\!\!\!\!\!\!
\Psi_a(E(S^{nm},\tau_i)E(S^{nm},\tau^i)\triangleright\phi)= \\ \nonumber  &&\!\!\!\!\!\! =(\frac {\theta}{8\pi^2})^2 \Psi_a(
B(S^{nm},\tau_i) B(S^{nm},\tau^i)\triangleright\phi)\ea finally
using (\ref{uu})
we get \ba \nonumber &&\!\!\!\!\!\!\!\!\!
\Psi_a(E(S^{nm},\tau_i)E(S^{nm},\tau^i)\triangleright\phi)= \\ &&\!\!\!\!\!\! \nonumber = (\frac {\theta}{4\pi^2})^2 \Psi_a(
{\rm tr}[U^{nm}\tau_i]{\rm tr}[U^{nm}\tau^i]\triangleright\phi + {\cal O}(\epsilon^4))= \\ && = (\frac {\theta}{4\pi^2})^2\delta_{n0}\delta_{m0} a_i a^i
\Psi_a(\phi)+{\cal O}(\epsilon^4),\ea where $a_i={\rm
Tr}(a\tau_i)$. In the previous equation we used the fact that for
$(n,m)\not=(0,0)$  $({\rm tr}[U^{nm}\tau_i]{\rm
tr}[U^{nm}\tau_i])=({\rm Tr}[\tau_i]{\rm Tr}[\tau_i])=0$. Putting
all this together we ntice that we can in this case take the limit $\epsilon\rightarrow 0$. The result is \be
\Psi_a(A(S)\triangleright\phi)=\frac {\theta}{4\pi^2}\gamma\sqrt{ a\cdot a}\
\Psi_a(\phi), \ee for all $\phi\in Cyl$. Our
generalized state $\Psi_a$ is an eigenstate of the area with an
eigenvalue that varies continuously as $a\in SU(2)$ varies.

\end{appendix}

\end{document}